\documentclass[preprint,5p,times,twocolumn]{elsarticle}
\journal{Nuclear Instruments and Methods A}

\usepackage{preamble}

\begin{document}

\begin{frontmatter}


\title{Characterization of the new Ultracold
  Neutron beamline at the LANL UCN facility}


\author[IU,LANL]{D.~K.-T.~Wong\corref{cor1}}
\ead{dkwong@iu.edu}
\author[LANL]{M.~T.~Hassan}
\author[LANL]{J.~F.~Burdine}
\author[UM]{T.~E.~Chupp}
\author[LANL]{S.~M.~Clayton}
\author[LANL,NCSU]{C. Cude-Woods}
\author[LANL]{S.~A.~Currie}
\author[LANL]{T.~M.~Ito\corref{cor1}}
\ead{ito@lanl.gov}
\author[IU]{C.-Y.~Liu}
\author[LANL]{M.~Makela}
\author[LANL]{C.~L.~Morris}
\author[LANL]{C.~M.~O'Shaughnessy}
\author[IU]{A.~Reid}
\author[UM,NW]{N.~Sachdeva}
\author[LANL]{W.~Uhrich}

\address[IU]{Indiana University, Bloomington, Indiana 47405, USA}

\address[LANL]{Los Alamos National Laboratory, Los Alamos, New Mexico 87545,
 USA}

\address[UM]{University of Michigan, Ann Arbor, Michigan, 48109, USA} 

\address[NCSU]{North Carolina State University, Raleigh, North Carolina,
27695, USA}

\address[NW]{Northwestern University, Evanston, Illinois, 60208, USA}

\cortext[cor1]{Corresponding authors}

\begin{abstract}

The neutron electric dipole moment (nEDM) experiment that is currently being developed at Los Alamos National Laboratory (LANL) will use ultracold neutrons (UCN) and Ramsey's method of separated oscillatory fields to search for a nEDM. In this paper, we present measurements of UCN storage and UCN transport performed during the commissioning of a new beamline at the LANL UCN source and demonstrate a sufficient number of stored polarized UCN to achieve a statistical uncertainty of $\delta d_n = 2\times 10^{-27}$~$e\cdot\text{cm}$ in 5 calendar years of running. We also present an analytical model describing data that provides a simple parameterization of the input UCN energy spectrum on the new beamline. 

\end{abstract}


\end{frontmatter}


\section{\label{sec:intro}Introduction}

Ultracold neutrons (UCN)~\cite{ignatovich1990physics,golubUCN} are neutrons of sufficiently low kinetic energy ($\lesssim$\qty{340}{\nano\eV}) that they can be confined in material or magnetic bottles. UCN are playing increasingly important roles in the studies of fundamental physical interactions (for recent reviews, see e.g. Refs.~\cite{Dubbers2011,Young2014}).

One particularly important class of experiments performed using UCN is the search for a nonzero electric dipole moment of the neutron (nEDM) (see e.g. Refs.~\cite{BAK06,SER15, ABE20}), which probes new sources of time reversal symmetry violation~\cite{POS05,ENG13,CHU19} and may give clues to the puzzle of the matter-antimatter asymmetry in the universe~\cite{Cir10,Mor13}. Modern nEDM experiments are performed almost exclusively using UCN (an alternative is proposed in Ref.~\cite{PIE13}). The current upper limit has been set by an experiment performed at the Paul Scherrer Institute~\cite{ABE20} and there are several efforts worldwide aimed at searching for the nEDM with improved sensitivity~\cite{Alarcon2022}.

\begin{figure}[htp]
    \centering
    \includegraphics[width=\columnwidth]{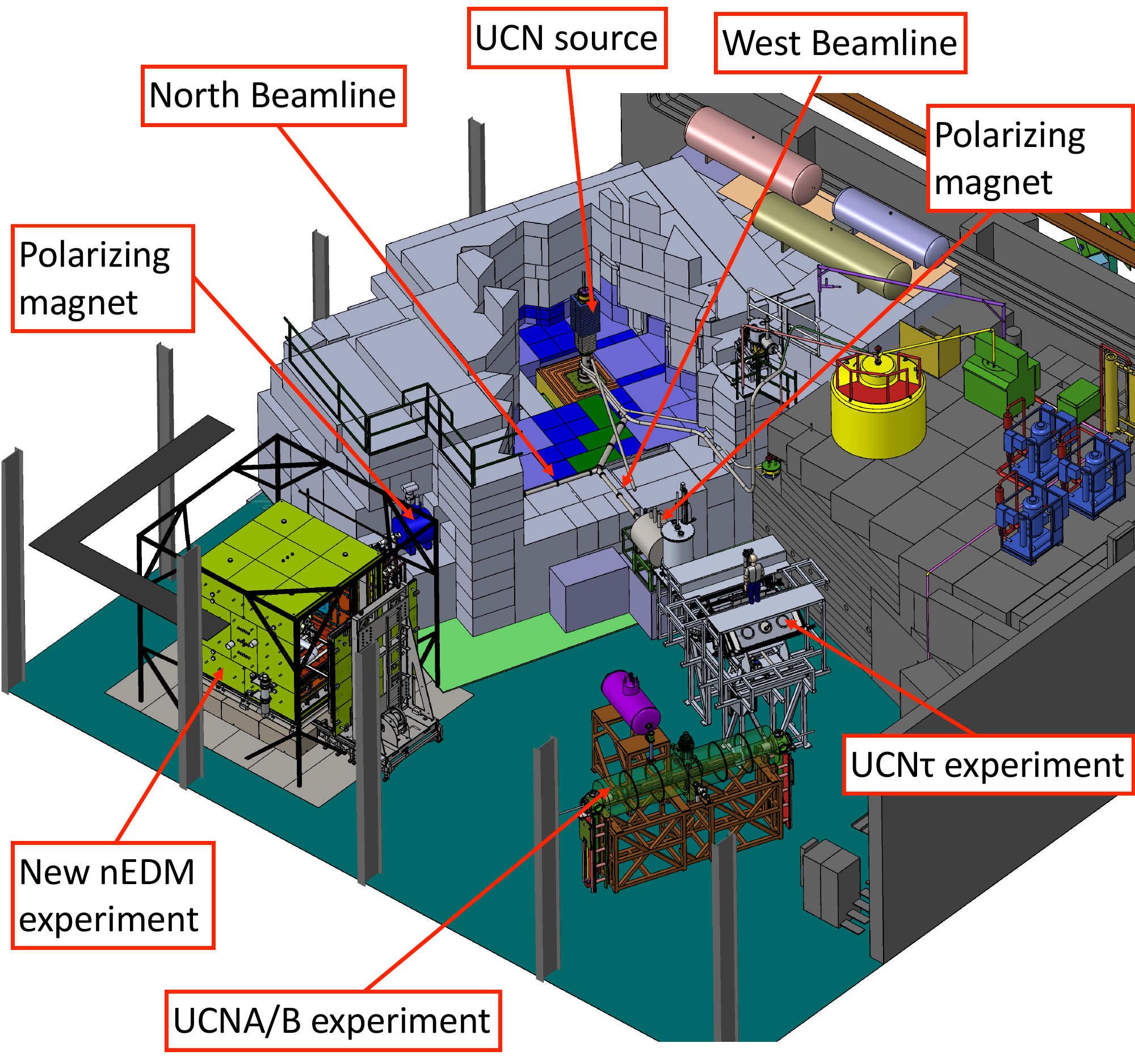}
    \caption{Schematic of the experimental area at the LANL UCN facility}
    \label{fig:AreaB_schematic}
\end{figure}

At the Los Alamos National Laboratory (LANL), a solid deuterium (SD$_2$) based superthermal UCN source coupled to a spallation target has been providing UCN to experiments for the last 20 years (see Fig.~\ref{fig:AreaB_schematic}). The UCN source has recently been upgraded to host a new nEDM experiment that will use Ramsey's method of separated oscillatory fields \cite{ramsey_molecular_1950} to search for the nEDM with an uncertainty goal of $\delta d_n = 2\times 10^{-27}$~$e\cdot\text{cm}$. In this upgrade, the cryogenic insert, which houses the SD$_2$ converter as well as the cold neutron moderator, was replaced with an improved design. The upgrade produced a factor of four increase in UCN density~\cite{ito_performance_2018}.

With the upgrade, a new UCN beamline (called the North Beamline) was constructed for the new nEDM experiment (see Fig.~\ref{fig:AreaB_schematic}). As part of the commissioning of the North Beamline and development of the new nEDM experiment, we performed a series of UCN transport and storage measurements using a prototype nEDM precession cell. 

In the envisioned nEDM experiment, a ``measurement cycle", which consists of the following steps, will be repeated many times over the course of the experiment: 1)~UCN are loaded into the precession cell(s), 2)~UCN are allowed to precess freely for a duration $T_\text{fp}$, 3)~UCN are unloaded from the precession cell(s) to be detected and to have their final spin state analyzed. The statistical uncertainty of each such measurement is given by 
\begin{gather}
    \sigma_{d_n} = \frac{\hbar}{2\alpha E T_\text{fp} \sqrt{N}}\label{eq:figure_of_merit},
\end{gather}
where $\hbar$ is Planck’s constant, $\alpha$ is a factor describing the spin contrast of a Ramsey fringe, $N$ is the number of the detected UCN, and $E$ is the strength of the applied electric field. $N$ is affected by UCN density at the location of the experiment, a result of UCN source performance and UCN transport. Both $N$ and $T_\text{fp}$ depend on the UCN storage property of the precession cell ({\it i.e.} how the number of stored UCN depends on the time over which UCN are stored in the precession cell), which is in turn affected by the input UCN energy spectrum as well as the wall loss properties of the cell. The operating condition ({\it i.e.} how long to load the cell, how long to store the UCN, and how long to unload the cell, etc) is optimized to minimize the uncertainty of the experiment. Therefore our measurements were focused on characterizing the UCN density and energy spectrum at the location of the experiment, as well as the UCN storage property of the precession cell. 

In this paper, we describe the measurements performed, present the results and analysis, and demonstrate an analytical model to quantify the collected data. 

\section{\label{sec:experimentalSetup}Description of experimental setup}

\begin{figure*}[htp]
    \centering
    \includegraphics[width=14cm]{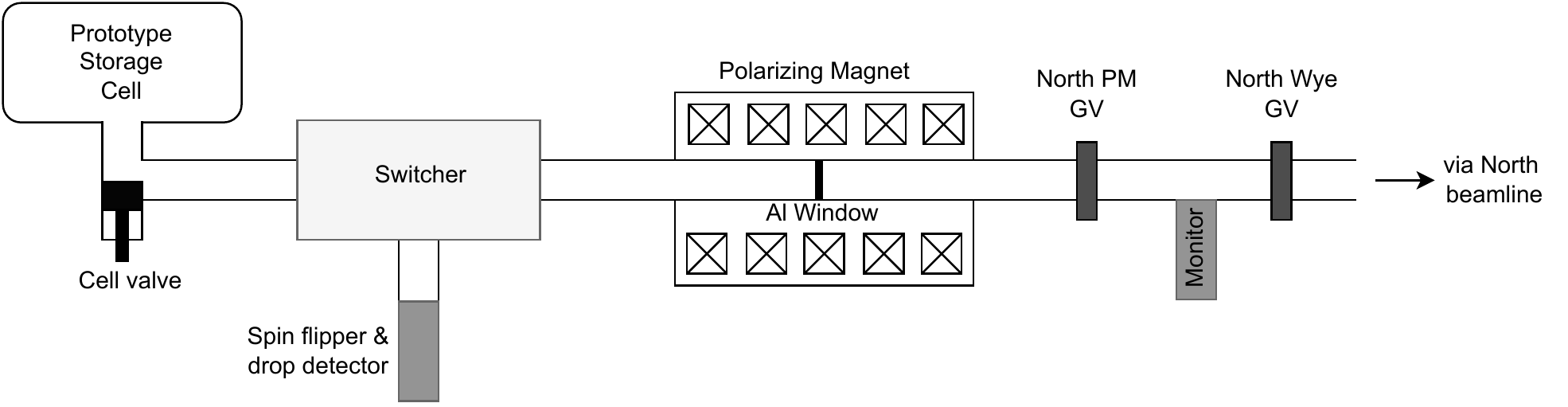}
    \caption{Layout of the North Beamline. Two types of switchers were tested, see Sec.~\ref{sec:experimentalSetup} for description}
    \label{fig:NorthBeamlineLayout}
\end{figure*}

The schematic diagram of the experimental setup mounted on the North Beamline is shown in Fig.~\ref{fig:NorthBeamlineLayout}. There were two gate valves (GV) located on the beamline, with a \qty{6.35}{\mm}-diameter ``monitor port'' located in between. UCN that traveled through the monitor port were counted with a $\ce{^{10}B}$ coated ZnS:Ag detector \cite{jeph_b10_2011}. The rate of UCN through the monitor port is proportional to the UCN density in the guides (in the case of an isotropic distribution, the rate $R$ through the monitor port hole with an area $A$ and the UCN density $\rho$ are related via  $R = \frac{1}{4} \bar{v} \rho A$, where $\bar{v}$ is the average UCN speed, {\it i.e.} $\bar{v} = \int vf(v)dv$, with $f$ describing the UCN velocity distribution). Hence, the rate of the monitor detector gives the approximate density of the UCN in the guide.

A 5-T, horizontal warm bore, superconducting magnet by American Magnetics Inc. was used as a polarizing magnet (PM). The strong magnetic field from the PM interacts with the neutron magnetic moment creating a potential given by $-\vec{\mu}_n\cdot\vec{B}$, where $\vec{\mu}_n$ is the neutron magnetic moment and $\vec{B}$ is the magnetic field. The PM field is effectively a UCN spin filter, acting as a potential barrier that rejects low-field seeking UCN below \qty{300}{\nano\eV} and a potential well that passes high-field seeking UCN. For the measurements described in this paper, we varied $ |\vec{B} |$ for the purpose of obtaining information on the UCN spectrum.

A 0.1-mm thick Al97 Mg3 alloy foil (similar in tensile strength to 6061-T6), termed the ``PM window", is located in the beamline at the center of the PM field region. This is used to separate the UCN source vacuum from the measurement apparatus vacuum, as the source vacuum can be routinely filled with D$_2$ gas ({\it e.g.} while D$_2$ is being drawn from a storage tank to be frozen or while SD$_2$ is being reconditioned). The magnetic potential of the PM overwhelms the neutron optical potential of the window (\qty{54}{\nano\eV} \cite{golubUCN}). The effect of the window on the transmission of UCN is discussed in detail in Sec.~\ref{sec:analysis}.

A UCN switcher was located downstream of the PM. The switcher directed UCN from the source to the cell (fill mode) or from the cell to a UCN spin analyzer and detector (counting mode). In the counting mode, UCN were sent to a ``drop-detector'' located below beam height. Two types of switchers were utilized for the measurements described in this paper. The first one was termed the ``prototype switcher,'' and the second, which will be used for the eventual LANL nEDM experiment, was termed the ``rotary switcher.'' 

\begin{figure*}[htp]
    \centering
    \includegraphics[width=7cm]{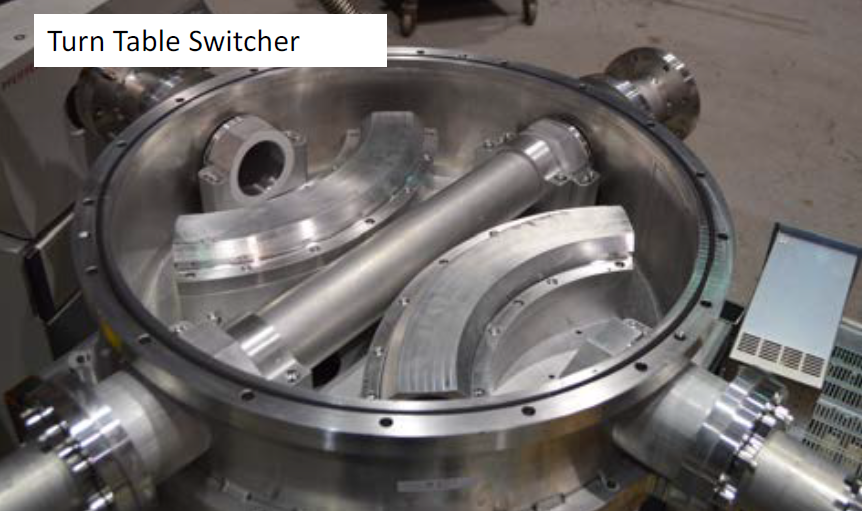}
    \hspace{1em}
    \includegraphics[width=\columnwidth]{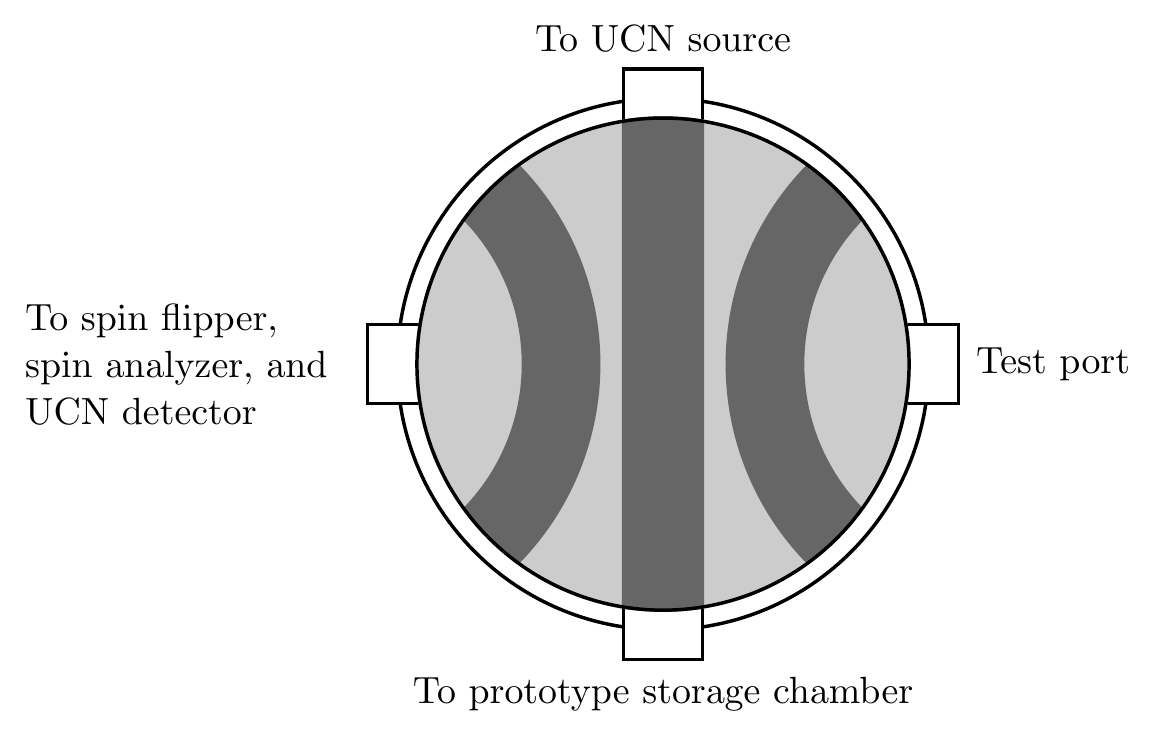}
    \caption{Left: A photograph of the rotary switcher. Right: Top-down schematic of the rotary switcher}\label{fig:NewSwitcher}
\end{figure*}

The prototype switcher contains a guide segment that can be set to a straight-through position to fill the cell, or angled downward to send UCN from the cell to the drop detector. The rotary switcher has a cylindrical housing and four evenly spaced guide ports. These guide ports could be connected internally by either a straight guide section or \ang{90} bend sections. All guide sections inside were mounted on a rotating turntable and actuated by a Parker CM232DX servo-motor. The drop detector was connected to the rotary switcher via a downward bend attached to one of the guide ports. An image of the rotary switcher as well as a schematic of the guide segments within it are shown in Fig.~\ref{fig:NewSwitcher}.

All guide sections downstream of the PM had a 50-${\mu}$m-thick electroless nickel phosphorus (NiP) coating. NiP has a high neutron optical potential of 213(5)~\unit{\nano\eV}, a low spin-flip probability per bounce of $3.6(5) \times 10^{-6}$, and a low UCN loss probability per bounce of $1.3(1) \times 10^{-4}$ \cite{tang_measurement_2016, pattie_jr_evaluation_2017}. To adiabatically transport neutron spins, the guides downstream of the PM were kept under a $\sim$1-mT magnetic field produced by coils.

On both the monitor and the drop-detector, $\ce{^{10}B}$ coated ZnS:Ag scintillator films were used for UCN detection \cite{jeph_b10_2011}. UCN were captured on the top layer of $\ce{^{10}B}$, and reaction products ($^4$He and/or $^7$Li) from the $\ce{^{10}B}(\text{n},\alpha)^7\text{Li}$ neutron capture reaction were detected in the ZnS:Ag layer. The resulting scintillation light was then detected by a photomultiplier tube (PMT). The photocathode diameters of the drop-detector and monitor detector PMTs were \qty{76.2}{\mm} and \qty{28.6}{\mm}, respectively.

\begin{figure}[htp]
    \centering
    \includegraphics[width=5cm]{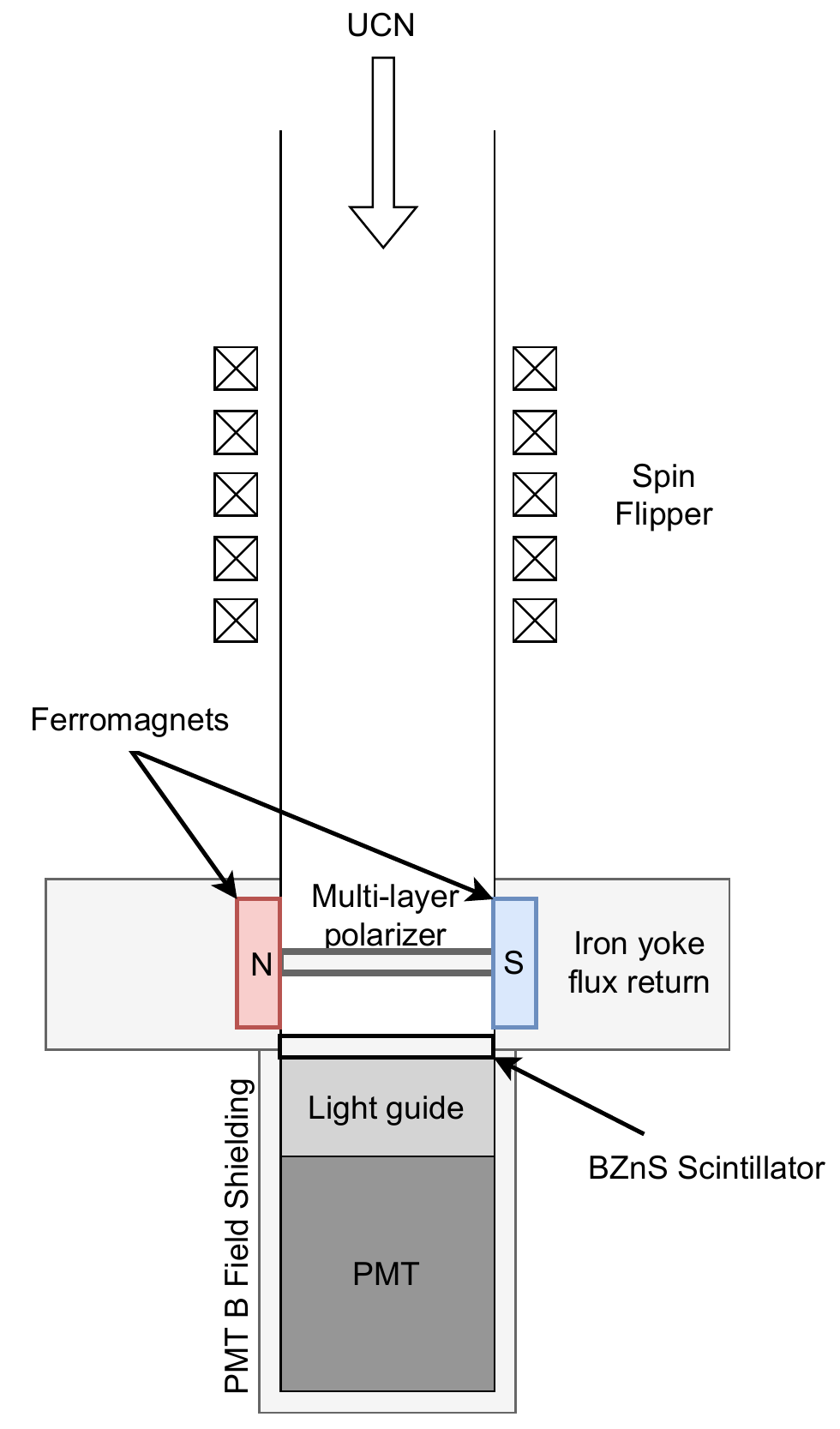}
    \caption{Schematic of the adiabatic fast passage (AFP) spin flipper, spin analyzer, and UCN drop detector}
    \label{fig:SpinAnalyzer}
\end{figure}

A neutron spin analyzer that consisted of a 10 layer polarizer made of iron and silicon was located immediately above the drop detector (see Ref.~\cite{ThorstenThesis} for layer structure details). When magnetized with a 10-mT field from permanent magnets, the multi-layer polarizer preferentially transmitted high-field seeking UCN and rejected low-field seeking UCN with an analyzing power of $99.3^{+0.7}_{-2.4}\%$ \cite{ThorstenThesis}. [Note that it is not clear if the neutron energy spectrum under which Ref.~\cite{ThorstenThesis} is consistent with our neutron spectrum. Reference~\cite{afach_device_2015} quotes an analyzing power of 95\% for a different but similar analyzing foil for an energy range of 90~neV to 330~neV, which includes the energy range of the UCN at the analyzing foil for the experiment reported in Sec.~\ref{sec:analysis}. The main conclusion of the results reported here does not depend on the exact value of the analyzing power. We use the value reported by Ref.~\cite{ThorstenThesis}.]   Additionally, an adiabatic fast passage (AFP) spin flipper coil, located upstream of the spin analyzer, was used to provide the option of spin flipping UCN \cite{holley_afp_2012}. A schematic of the analyzer system is illustrated in Fig.~\ref{fig:SpinAnalyzer}.

UCN were stored in a prototype nEDM precession chamber (hereafter simply referred to as ``the cell"). The cell was a cylinder with an inner diameter of \qty{50}{\cm} and a height of \qty{10}{\cm}, giving a total volume of \qty{19.6}{\liter}. Several different configurations for the cell were used in the measurements for this paper and are discussed in further detail in Sec.~\ref{subsec:holdingTimeMeasurement}. The bottom of the cell was located approximately \qty{10}{\cm} above beamline guide height.

\begin{figure}[htp]
    \centering
    \includegraphics[width=\columnwidth]{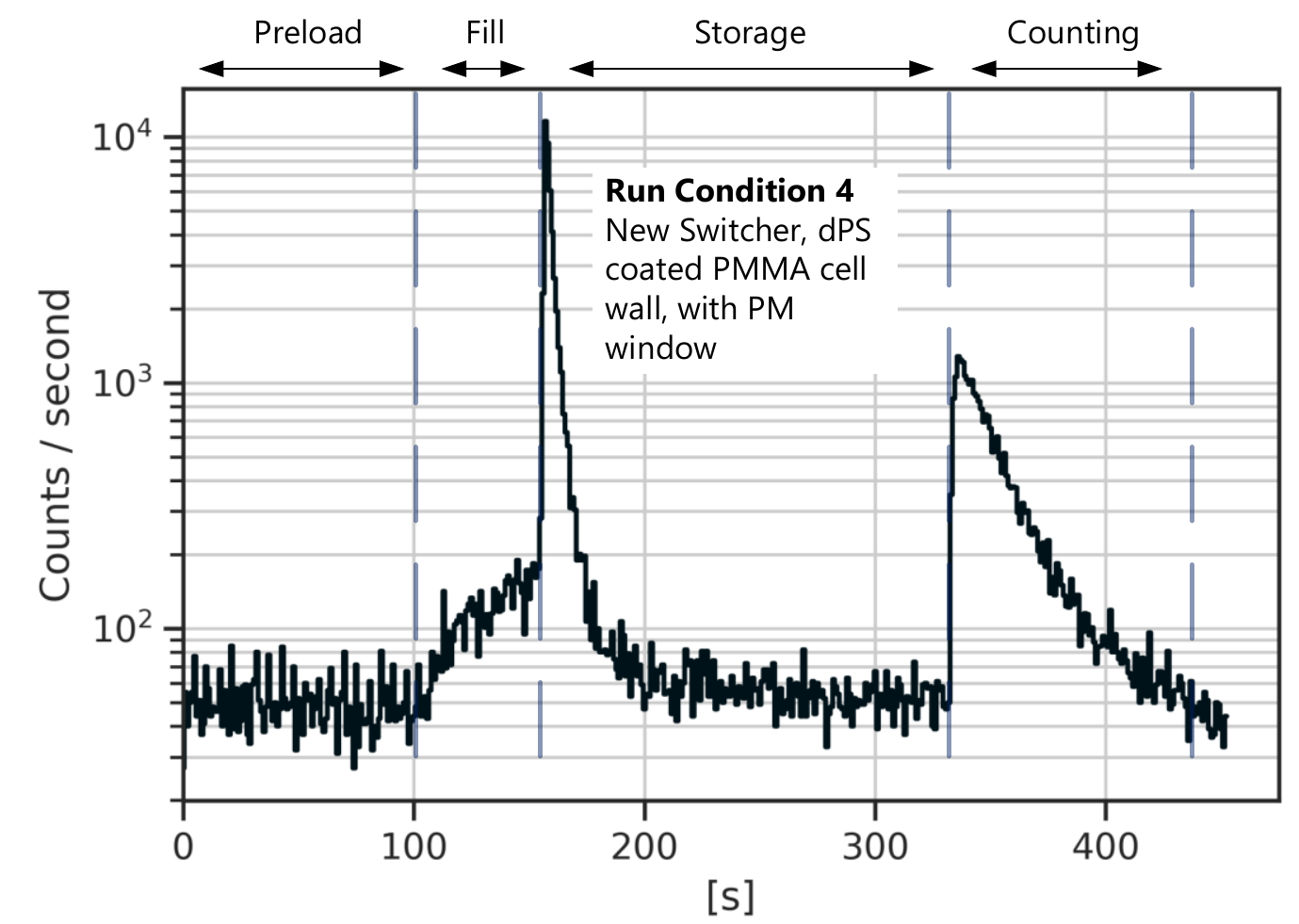}
    \caption{UCN count rate during the ``fill-and-dump" measurement sequence for one run. Preload: \qtyrange{0}{100}{\s}. Filling period: \qtyrange{100}{150}{\s}. Holding period: \qtyrange{150}{330}{\s}. Counting period: \qtyrange{330}{430}{\s}.}
    \label{fig:timeSpectrum}
\end{figure}

\begin{figure}[htp]
    \centering
    \includegraphics[width=\columnwidth]{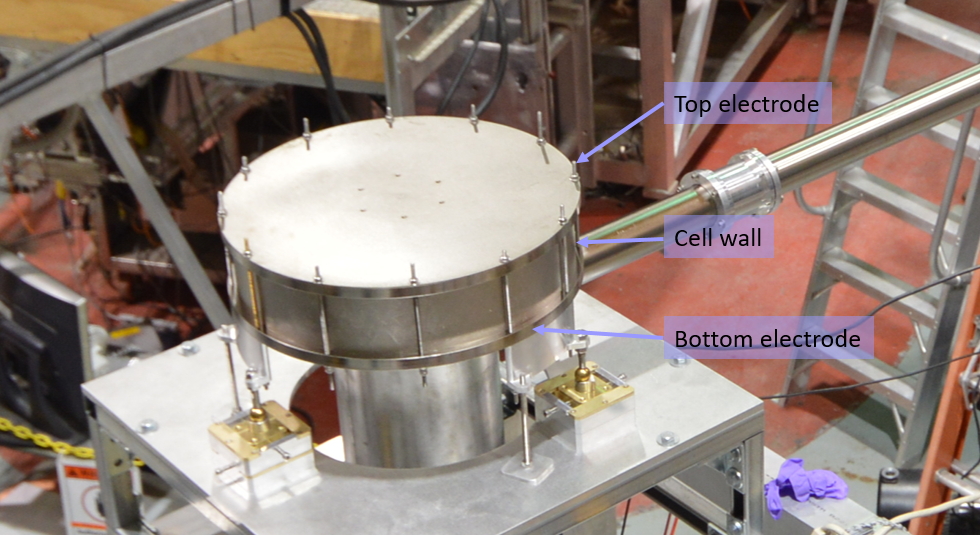}
    \caption{A photograph of a prototype precession cell (pictured here with NiP-coated Al cell walls)}
    \label{fig:dummyCell}
\end{figure}

\section{\label{sec:measurement}Measurements}

\subsection{\label{subsec:holdingTimeMeasurement}Measurement of UCN storage time}

A UCN storage lifetime measurement was performed using a ``fill-and-dump" method. The steps were: (1)~The North Wye GV was opened while the North PM GV was kept closed for \qty{100}{\s} to ``preload" the UCN to build up UCN density behind the GV. (2)~The North PM GV was opened, allowing UCN to pass through the PM and the switcher into the cell. This filling period was \qty{50}{\s}. (3)~The cell valve and North PM GV were closed to stop the flow of UCN into the precession cell. (4)~The cell valve was kept closed for some ``holding time" to store UCN. (5)~The switcher moved from the fill mode to count mode in preparation to empty the cell. (6)~The cell valve was opened to allow UCN to flow to the drop detector. This counting period lasted for \qty{100}{\s}. Note that for these of measurements, the UCN final spin was not measured. See Fig.~\ref{fig:timeSpectrum} for the drop detector UCN count rate for one such run.

Two different types of cell side walls were tested: (i)~NiP coated Al and (ii)~deuterated polystyrene (dPS) coated poly(methyl methacrylate) (PMMA). In both cases, the top and bottom surfaces of the cell were NiP-coated Al.  A photograph of the prototype cell is shown in Fig.~\ref{fig:dummyCell}. In the final nEDM experiment the dPS-coated PMMA cell side wall will be used in combination with top and bottom plates made of Al coated with a diamond-like carbon (DLC), which will be used as ground and high voltage electrodes. We used NiP coating, instead of DLC, for the reason that it was easier to obtain. However, given that DLC has a higher neutron optical potential~\cite{Atchison2006} than NiP and also given that the loss parameter we obtained for NiP was similar to that measured for DLC (see Sec.~\ref{sec:analysis}), a cell made of a dPS-coated PMMA side wall and the NiP coated plates provides conditions sufficiently close to the final nEDM experiment. To further approximate the condition of the final nEDM experiment, the top and bottom plates have a groove with rounded edges where the end of the insulating wall meets them, a feature needed to avoid enhanced electric fields (see, e.g. Sec. 3.3 of Ref.~\cite{KHR97}).  

The seals between the side wall and the top and bottom plates were provided by an O-ring, located inside the groove mentioned above, minimizing the exposure to UCN. To ensure that the storage time was not affected by upscattering of UCN from residual gas, we maintained a storage volume vacuum of $10^{-5}$~Torr or better. A high voltage was not applied to the high voltage electrode (although it was possible when the side wall was the one made of dPS-coated PMMA), as it was not essential for the purpose of the measurement. As a result, the cell was placed in the atmosphere (as shown in Fig.~\ref{fig:dummyCell}) as opposed to in an insulating vacuum. 

Fill-and-dump holding time scans were repeated for holding times of 30 to 300 seconds. Holding time scans were repeated with PM strengths of \qtylist{0; 1.25; 2.5; 3.75; 5}{\tesla}. An overview of UCN hold-time run conditions is summarized in Tab.~\ref{tb:runconditions}. For run condition 1, the UCN source was run continuously and the GV was simply opened and closed between fill-and-dump measurements. To avoid delivering beam to the target unnecessarily, run conditions 2 -- 4 switched to a mode where the beam was on only during preload and filling periods.

\begin{table*}
\centering
\caption{\label{tb:runconditions}Switcher, cell wall, and PM window conditions for measurements} 
\begin{tabular}{lll}
\toprule
\multicolumn{1}{l}{$\#$} & \multicolumn{1}{l}{Run Condition}  & \multicolumn{1}{l}{Storage Curve} \\
\midrule
 1 & Prototype switcher, NiP coated Al cell wall, no  PM window & Fig. \ref{fig:runCondition1_fit} \\  
 2 & Prototype switcher, NiP coated Al cell wall, with PM window & Fig. \ref{fig:runCondition2_fit} \\
 3 & Rotary switcher, NiP coated Al cell wall, with PM window & Fig. \ref{fig:runCondition3_fit} \\
 4 & Rotary switcher, dPS coated PMMA cell wall, with PM window & Fig. \ref{fig:runCondition4_fit} \\
\bottomrule
\end{tabular}
\end{table*}

\subsection{\label{subsec:PMrampMeasurement}Measurement of spin dependent UCN transmission rate as a function of PM field}

Direct ``flow-through" measurements of UCN transmission through the PM were taken to provide information on the incoming UCN energy spectrum. During these measurements, the magnetic field of the PM was slowly decreased from \qtyrange{5}{0}{\tesla} or increased from \qtyrange{0}{5}{\tesla}, depending on the initial state of the PM. \qty{44}{\minute} are needed to ramp between 0~T and 4.11~T, and \qty{28}{\minute} to ramp between 4.11~T and 5.0~T, a total of \qty{72}{\minute}. UCN were transported from a continuously operated source to the drop detector, while the AFP spin flipper state was toggled every \qty{67}{\s}. The period of 67~s (= 0.015~Hz) was chosen to ensure sufficiently frequent toggling between the two spin states while providing a long enough period for each segment of measurement needed for accurate analysis. These measurements were performed with the Al window installed in the PM region. Analysis of this data is presented in Sec.~\ref{subsec:PMramp}.

\begin{figure}[htp]
    \centering
    \includegraphics[width=\columnwidth]{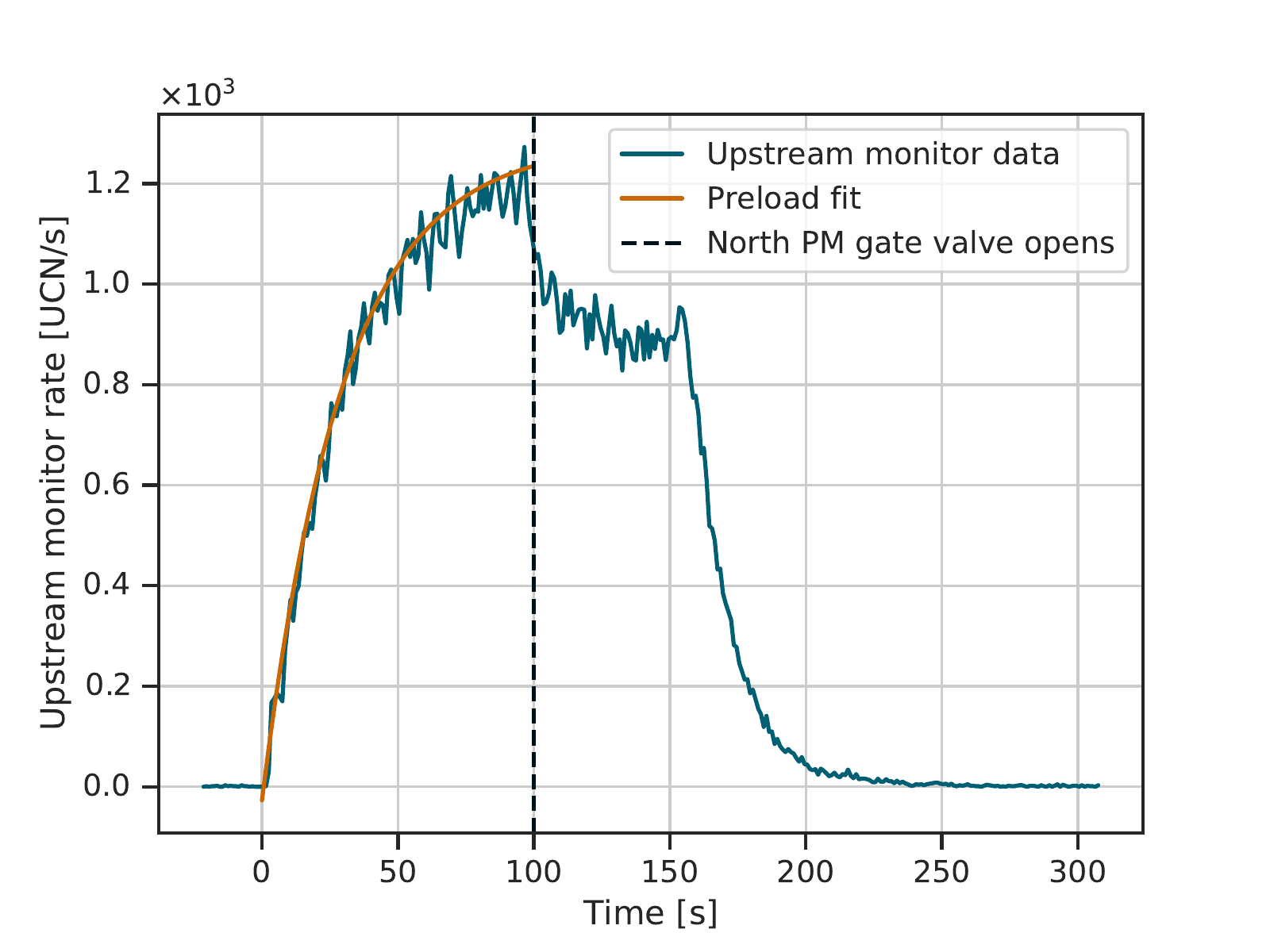}
    \caption{An illustration of a preload fit used for normalization under run conditions 2 -- 4, as described in section \ref{subsec:dataProcessing}. Data in this figure is obtained from an upstream monitor located on the ``West Beamline''. UCN density is built up behind a stainless steel gate valve for \qty{100}{\s}. The gate valve is then opened to allow UCN to fill the precession cell on the north beamline, causing a dip on the upstream monitor rate}
    \label{fig:preloadFit}
\end{figure}

\section{\label{sec:analysis}Data Analysis and Modeling}

\subsection{\label{subsec:dataProcessing}Data processing and normalization}

In order to generate a reliable UCN storage time curve, it is necessary to subtract background and normalize UCN counts to correct for variations in UCN source intensity.

A monitor upstream of the PM located on the ``West Beamline" (see Fig.~\ref{fig:AreaB_schematic}) was used to normalize the number of UCN measured by the drop detector during the counting period. This monitor was chosen for normalization because the UCN source performance was characterized via West Beamline monitor before the addition of the North beamline. The West beamline monitor measured a UCN count rate of $\sim2000$~UCN/s (North PM GV closed) when beam and source parameters were optimally tuned. This monitor rate corresponded to a density of 184(32) UCN$/\text{cm}^3$ at the exit of the biological shield (see Ref.~\cite{ito_performance_2018}).

The UCN rate measured in the West Beamline monitor as a function of time during the preload, $R(t)$, was fit to the form of an exponential, $R(t) = R_0( 1-\exp[-(t-t_0) /\tau ] )$, to determine free parameters $R_0,\,t_0,\text{ and } \tau$ for run conditions 2 -- 4 in Tab. \ref{tb:runconditions}. The scaling factor to obtain a normalized UCN count is then given by $2000 / R(t=100\;{\rm s})$, the rate at the end of the preload period. An example of a preload fit utilized for normalization is illustrated in Fig.~\ref{fig:preloadFit}. For run condition 1, $R$ was taken to be the average value of the West Beamline monitor rate during the preload period because the beam was not turned off during the measurement run. The PMT background rate for each data set was fitted and subtracted from the UCN counting period.

During analysis, it was discovered that some portion of UCN storage runs had a small leakage of UCN during the storage period, a result of the cell valve not fully closing. Typically, in a leak-less run, UCN remaining in the guide system after the filling period drained into the drop-detector in the first $\sim$~\qty{20}{\s} of the holding period. This ``guide-dump'' count of UCN could be fit with a single decaying exponential and typically exhibited a decay time on the order of one or two seconds. However, with a leak, a double exponential function was observed with a secondary decay on the order of $\sim$~\qty{50}{\s}. 

\begin{figure}[htp]
    \centering
    \includegraphics[width=\columnwidth]{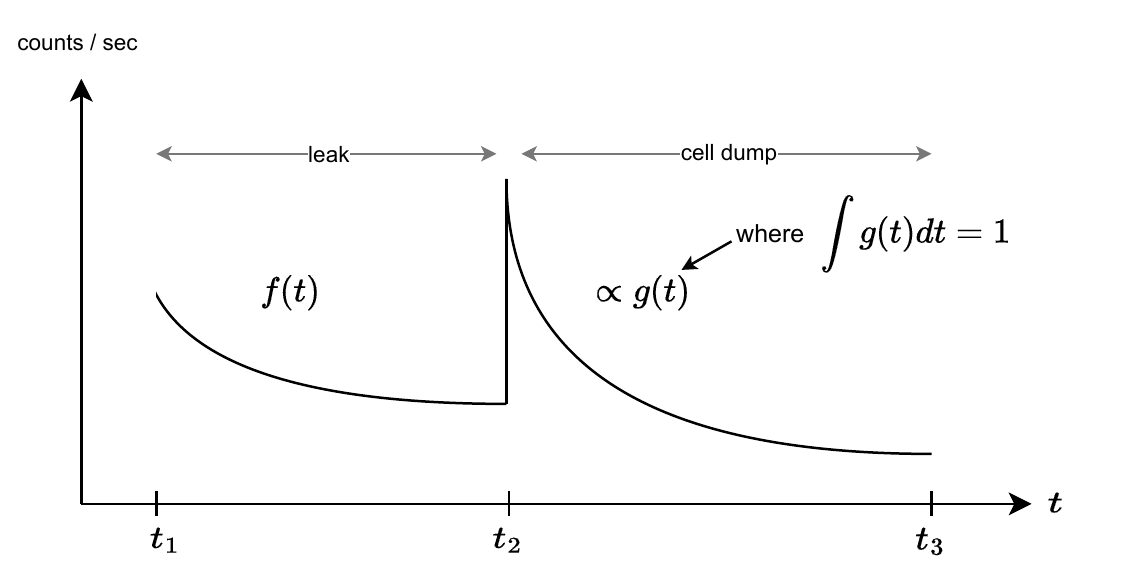}
    \caption{Visualization of parameters used for leak correction of UCN storage in Eq.~ (\ref{eq:leakCorrection}), as referred to in section \ref{subsec:dataProcessing}}
    \label{fig:leakCorrection}
\end{figure}

The leak-corrected count of UCN, $N_\text{cc}$, was obtained from the uncorrected count of UCN, $N_\text{c}$, with

\begin{gather}
    N_\text{cc} = N_\text{c} + \int^{t_2}_{t_1} dt \int^{t_3}_{t_2} dt'\, f(t)g(t') e^{-(t'-t)/\tau}\label{eq:leakCorrection}
\end{gather}
where, as illustrated in Fig.~\ref{fig:leakCorrection}, $t_1$ is the start of the holding period, $t_2$ is the start of the counting period, and $t_3$ is the end of the counting period. $N_\text{c}$ is obtained from the UCN counted during the counting period with background subtracted. The number of leaked UCN at a given time, $f(t)$, was obtained by fitting the guide dump on leak-free runs, and then subtracting the fit from runs with leaks. The rate of UCN detected at time $t$ during the counting period from the cell dump, $g(t)$, is normalized such that $\int g(t) dt = 1$.  The exponential term approximates the decay of stored UCN, where $\tau$ is the effective decay constant determined by fitting to the uncorrected storage curve.

\begin{figure}[htp]
    \centering
    \includegraphics[width=\columnwidth]{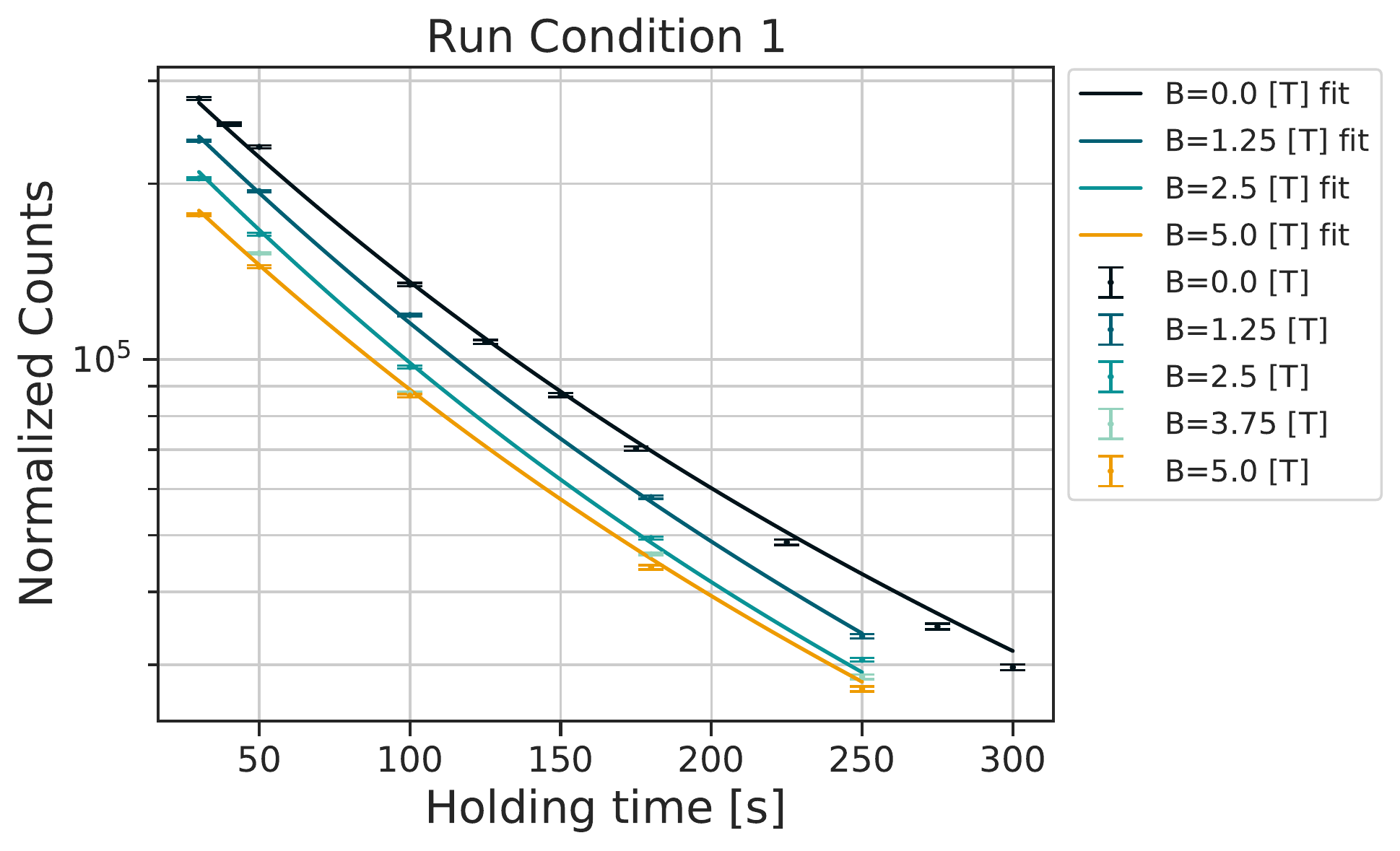}
    \caption{UCN storage measurement run condition 1: Prototype switcher, NiP coated Al cell wall, no PM window. Points with error bars represent data and curves represent fits. The $B=3.75$~T fit is suppressed due to overlap with the $B=5$~T curve. Fit results are described in Tab.~\ref{tb:fitResults}.}
    \label{fig:runCondition1_fit}
\end{figure}

\begin{figure}[htp]
    \centering
    \includegraphics[width=\columnwidth]{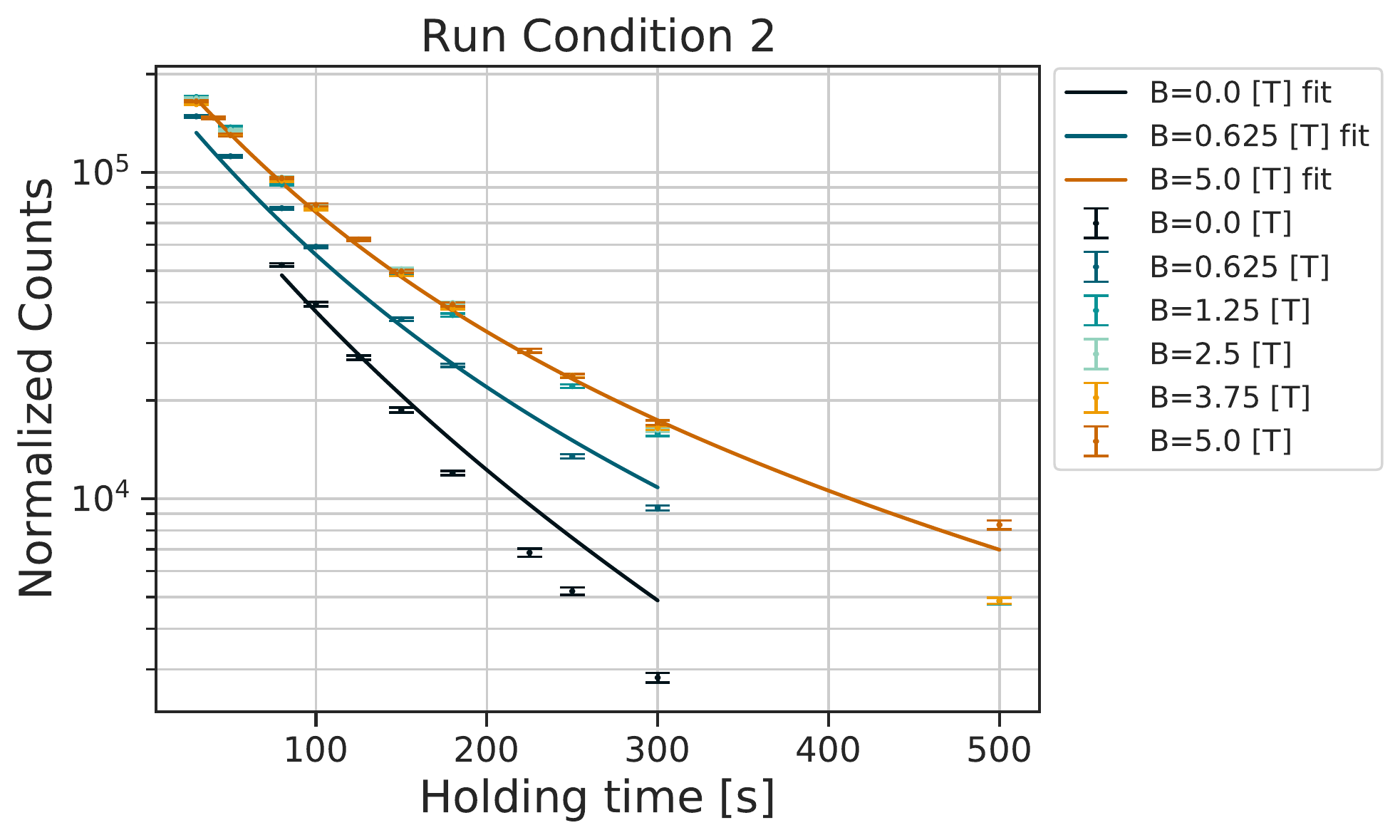}
    \caption{UCN storage measurement run condition 2: Prototype switcher, NiP coated Al cell wall, with PM Window. Points with error bars represent data and curves represent fits. The $B=1.25$~T, $B=2.5$~T and $B=3.75$~T fits are suppressed due to overlap with the $B=5$~T curve. Fit results are described in Tab.~\ref{tb:fitResults}.}
    \label{fig:runCondition2_fit}
\end{figure}

\begin{figure}[htp]
    \centering
    \includegraphics[width=\columnwidth]{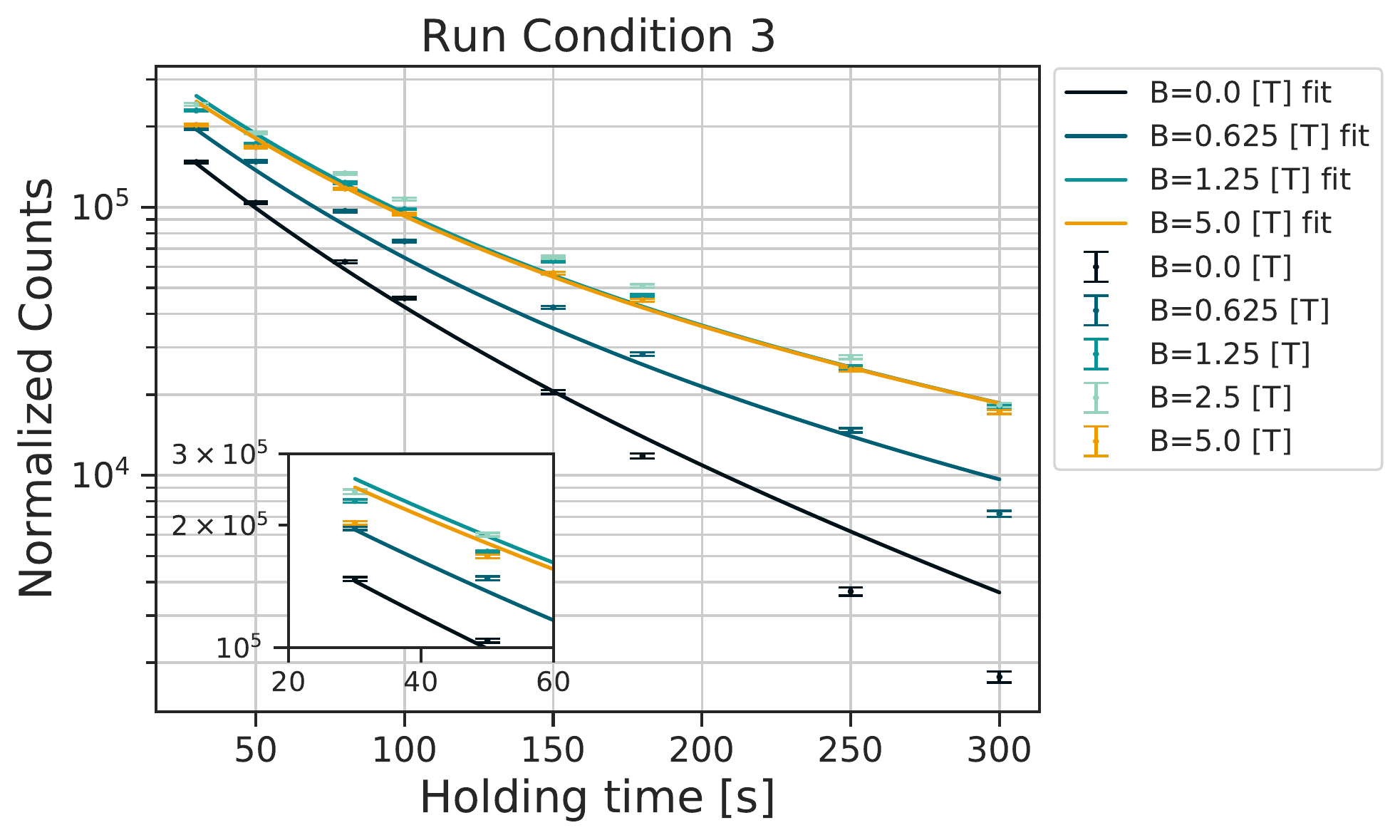}
    \caption{UCN storage measurement run condition 3: Rotary Switcher, NiP coated Al cell wall, with PM window. Points with error bars represent data and curves represent fits. The $B=2.5$~T fit is suppressed due to overlap with the $B=5$~T curve. Fit results are described in Tab.~\ref{tb:fitResults}.}
    \label{fig:runCondition3_fit}
\end{figure}

\begin{figure}[htp]
    \centering
    \includegraphics[width=\columnwidth]{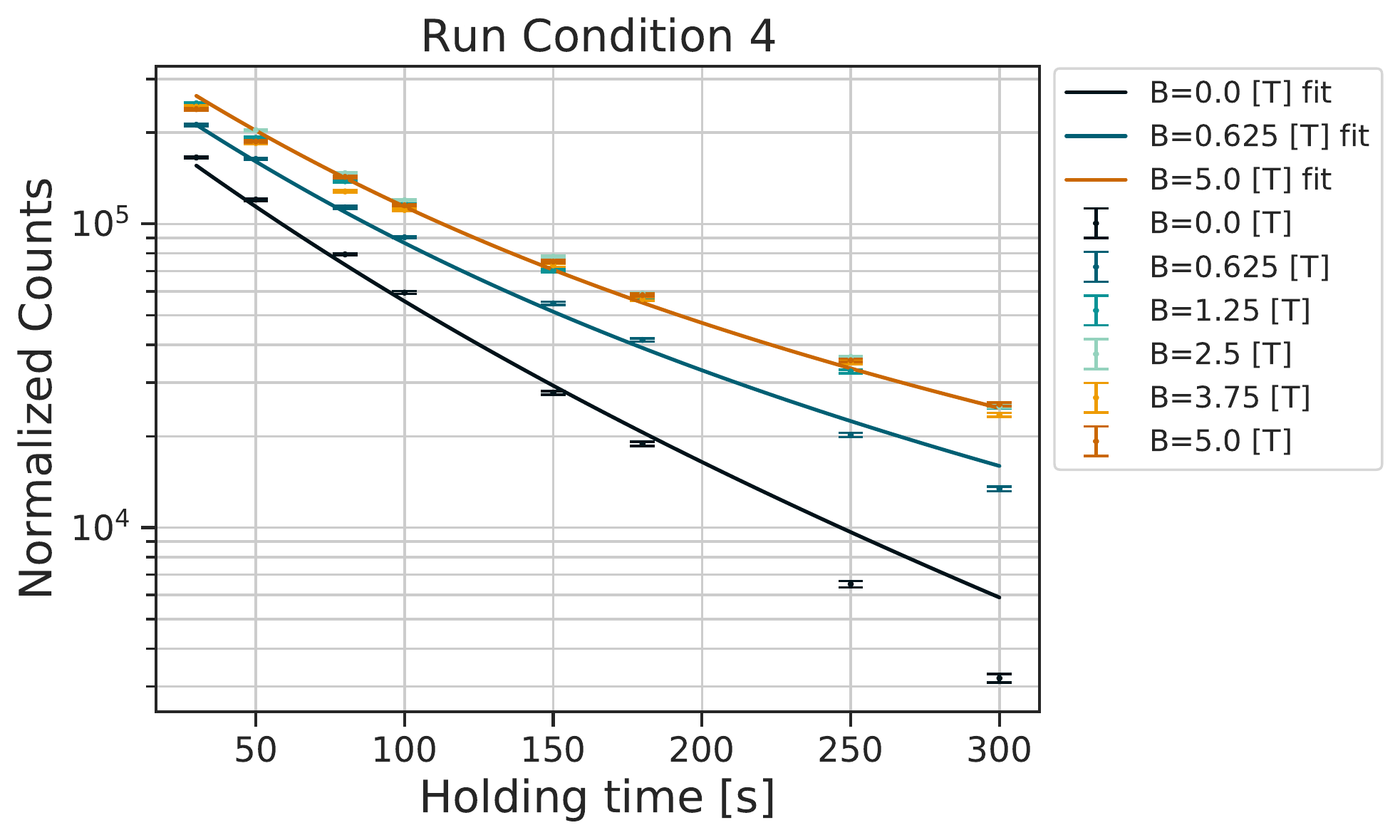}
    \caption{UCN storage measurement run condition 4: Rotary Switcher, dPS coated PMMA cell wall, with PM window. Points with error bars represent data and curves represent fits. The $B=1.25$~T, $B=2.5$~T and $B=3.75$~T fits are suppressed due to overlap with the $B=5$~T curve. Fit results are described in Tab.~\ref{tb:fitResults}}
    \label{fig:runCondition4_fit}
\end{figure}

The size of the leak correction varied. It was smaller for shorter holding time runs ($ < 2\%$ of counted UCN for storage times less than \qty{50}{\s}), and was larger for longer hold runs (up to 10\% of counted UCN for storage times of \qty{300}{\s}). The uncertainty of the leak correction was less than 5\% of the leak correction itself, as determined by taking into consideration all fitting and statistical uncertainties, making a negligible contribution to total measurement uncertainty. The normalized UCN counts thus obtained are plotted in Fig.~\ref{fig:runCondition1_fit} -- \ref{fig:runCondition4_fit}.

\subsection{\label{subsec:storageCurves}UCN storage curves in prototype precession cells}

In this section we introduce an analytical model to fit UCN lifetime storage curves in the cell, with parameters that describe properties of the UCN storage vessel, physical parameters of the neutron velocity spectrum upstream of the PM, and the effect of the PM on the spectrum. This model can be used to quickly obtain physical information from UCN storage curves without the need to create a comprehensive Monte Carlo simulation of a system, which can be labor intensive and computationally costly.

It is a common practice to fit UCN storage lifetime measurements to single-exponential curves, which is only correct for a population of UCN with a single velocity and a single angle of incidence. However, both the velocity and angle have continuous distributions. Double exponential curves are often used to very approximately represent decay curves made of a two-component population including a super-barrier fraction. In cases where the UCN velocity spectrum affects the performance of the experiment, such as an nEDM experiment, a more sophisticated approach is in order. 

Let $dN_0/dv$ be the UCN velocity spectrum upstream of the PM, and $dN_1/dv$ be the UCN velocity spectrum downstream of the PM. As the starting point, for simplicity, assume the following functional forms:

\begin{align}
    \frac{dN_0}{dv} &\propto v^\gamma \label{eq:dN0dv} \\ 
    \frac{dN_1}{dv} &= \frac{1}{2} \frac{dN_0}{dv} \left( f_h + f_\ell \right)\label{eq:dN1dv_noBeta}
\end{align}
where $\gamma$ is a free parameter of the model. Equation~(\ref{eq:dN0dv}) is a functional form for the velocity distribution of the UCN as a result of the transport from the source to the location of the experiment, as utilized in Ref.~\cite{ito_performance_2018}. $f_h$ and $f_\ell$ are the fraction of high-field seekers and low-field seekers transmitted through the PM, respectively. The normalization of $f_h$ and $f_\ell$ is chosen such that $f_h=1$ and $f_\ell=1$ corresponds to the completely unpolarized case.  

In the situation with no window present in the center of the PM, the PM passes all the high field seeking neutrons with $v_\parallel>0$, where $v_\parallel$ is the longitudinal velocity along the axis of the UCN guide. The PM passes low field seeking neutrons for $v_\parallel>0$ only if
$\frac{1}{2}m_n \left( v_\parallel \right)^2 > \mu_n B$, where $\mu_n$ is the neutron magnetic moment ($\mu_n=$~\qty
{60}{\nano\eV\per\tesla}) \cite{codata_2018}. Let $v^c_\parallel$ be the critical longitudinal velocity that fulfills this requirement, given by $v^c_\parallel = \sqrt{2\mu_n B/ m_n}$.

There is another factor that contributes to the transmission of UCN through the PM. If there is a spin-depolarizing region upstream of the PM (for example, we have gate valves made of stainless steel) and if the population of low-field seekers exceeds that of high-field seekers in the region upstream of the PM, which in turn is caused by low-field seekers being reflected by the PM potential (resulting in low $f_\ell$), there is a chance for low-field seekers to be spin-flipped and pass through the PM as high-field seekers. We can rewrite Eq.~(\ref{eq:dN1dv_noBeta}) as

\begin{align}
    \frac{dN_1}{dv} = \frac{1}{2} \frac{dN_0}{dv} \left[ f_h + f_\ell + \beta \left( f_h - f_\ell \right)  \right]
    \label{eq:dN1dv}
\end{align} 
where $\beta$ is a free parameter (constrained to the range $[0,1]$) that describes this spin-flipping effect.

We now describe expressions for $f_h$ and $f_\ell$. For the no-window case we set $f_h = 1/2$, a value corresponding to the situation in which half of the UCN are directed upstream and half are directed downstream.  Strictly speaking, this holds only when the system has no loss. This condition is reasonably well met for the measurement of storage times (described in Sec.~\ref{subsec:holdingTimeMeasurement}) but is not at all met for the measurement of spin dependent UCN transmission (described in Sec.~\ref{subsec:PMrampMeasurement}) because in the latter case, all the transmitted neutrons are eventually detected. Nevertheless, this formalism is still valid because the backward flowing neutrons do not come into the picture and because there is an overall normalization constant $C$. It should be considered to be a choice of convention to set $f_h= 1/2 $ for no window case.

$f_\ell$ is a function of velocity that is dependent on the angular distribution of the UCN upstream of the PM and the strength of the PM field. For an isotropic angular distribution, the fraction of low field seekers allowed through the PM is

\begin{align}\label{eq:fL_isotropic}
    f_\ell &= \frac{2\pi \int^{\theta_c}_{0}\: d\theta \sin\theta}{4\pi} = \frac{1}{2}\left(1 - \cos\theta_c\right) \\
    &= \frac{1}{2}\left( 1 - \frac{v^c_\parallel}{v} \right) \hspace{10pt} \text{ for } v>v^c_\parallel
\end{align}
The limits of integration in the numerator were determined using the requirement $\cos\theta > v^\parallel_c / v$ for low-field seeking neutrons.

As UCN are transported from the source to the experiment through the guide system, the angular distribution becomes more forward directed. For a completely forward directed population of low-field seekers, we have $f_\ell = 1/2 \text{ for } v>v^c_\parallel$ and $f_\ell = 0$ otherwise. The angular distribution is likely to be somewhere in between. We write a more general expression for $f_\ell$ with a free parameter characterizing the angular distribution of low field seeking neutrons

\begin{gather}
    f_\ell=\frac{1}{2} \left( 1 - k\left(\frac{v^c_\parallel}{v}\right) \right)  \hspace{10pt} \text{ for } v>v^c_\parallel\label{eq:fL_k}
\end{gather}
where as $k$ approaches 0, the UCN velocity distribution becomes forward directed, and as $k$ approaches 1 the angular distribution becomes isotropic.

We now describe a set of equations that can be used to fit UCN storage curves in a precession cell

\begin{align}
    N(t) &= \int_{v_\text{low}}^{v_\text{high}} dv \frac{dN_1}{dv}e^{-\frac{t}{\tau(v',\bar{\mu})}}\label{eq:holdingCurve_noWindowStart} \\
    \frac{1}{\tau(v',\bar{\mu})} &=\frac{1}{\tau_\beta}+\frac{Av'}{4\text{ Vol}}\bar{\mu} \label{eq:1overTau}\\
     \bar{\mu} &= 2\mu \left[ \frac{V^{\text{cell}}}{K(v')} \sin^{-1}\left( \frac{K(v')}{V^{\text{cell}}}\right)^{1/2} -\left( \frac{V^{\text{cell}}}{K(v')} - 1 \right)^{1/2} \right]\label{eq:lossPerBounce}\\
    \frac{dN_1}{dv} &= \frac{C}{2} v^\gamma \left[ \frac{1}{2} + f_\ell + \beta \left( \frac{1}{2} - f_\ell \right)  \right] \label{eq:dN1dv_no_window}\\
    f_\ell &=\begin{cases}
        \frac{1}{2} \left( 1 - k\left(\frac{v^c_\parallel}{v}\right) \right) \text{ for } v>v^c_\parallel \text{ where } v^c_\parallel = \sqrt{2\mu_N B/ m_N} \\
        0 \hspace{53pt}\text{otherwise}
        \end{cases}\label{eq:holdingCurve_noWindowEnd}
\end{align}
with free parameters $C$, $\mu$ (velocity-independent loss per bounce within the precession cell), $\beta$ (upstream depolarization probability), $\gamma$ (velocity distribution), and $k$ (angular distribution). $N$ (without any subscripts) refers to the number of detected neutrons. 

The lower integration limit of Eq.~(\ref{eq:holdingCurve_noWindowStart}), $v_\text{low}$, is obtained from the gravitational potential $U$, the minimum energy of UCN required to enter a precession cell located above beam height.  The velocity of UCN that have entered the cell, $v'$, is related to $v$ by $v' = \sqrt{v^2 - 2U/m_n}$. The term in the upper integration limit, $v_\text{high}$, is the velocity equivalent of $\min( V^{\text{cell} } + U,\, V^{\text{guide} })$, which is the maximum UCN energy from the input spectrum that is containable by either the transport guide (with neutron optical potential $V^{\text{guide}}$) or the precession cell (with neutron optical potential $V^{\text{cell} }$). Equation~(\ref{eq:1overTau}) describes the decay time of a stored UCN, which consists of the the free neutron lifetime $\tau_\beta$ with a velocity-dependent loss per bounce within the precession cell, where we assume that neutron velocity within the cell is isotropic. $A$ is the inner surface area of the precession cell, and Vol is the volume of the precession cell. Equation~(\ref{eq:lossPerBounce}), which describes loss per bounce, is taken from Eq.~2.70 in Ref.~\cite{golubUCN}, where $K(v')$ is the kinetic energy of the neutron in the cell. We assume that the UCN spectrum is not affected by the transport between the exit of the PM and the UCN precession cell. Therefore $dN_1/dv$ also represents the input UCN velocity spectrum of UCN in the guide that arrive at the cell.

The holding curve formalism described by Eqs.~(\ref{eq:holdingCurve_noWindowStart}) -- (\ref{eq:holdingCurve_noWindowEnd}) can also be modified to account for the presence of the Al window in the PM region. We maintain the base form of Eq.~(\ref{eq:dN1dv}), but now must alter our expressions for $f_\ell$ and $f_h$ to account for the step potential represented by the window, which for Al we approximate to be $V^\text{Al}=54\text{ neV}$ \cite{golubUCN}. Low-field seekers must now have a high enough velocity to overcome both the PM field potential and the Al window step potential. 

High-field seekers must also have a high enough velocity to overcome the window step potential, but receive the benefit of being accelerated by the PM field. When the energy increase from the PM field is greater than the neutron optical potential of the window, high-field seekers of any velocity are transmitted through the window.

To describe UCN storage curves with an Al window in the PM region, Eqs.~(\ref{eq:dN1dv_no_window}) and (\ref{eq:holdingCurve_noWindowEnd}) become

\begin{align}
     \frac{dN_1}{dv} &= \frac{C}{2}v^{\gamma} \left[ f_h + f_\ell + \beta \left( f_h - f_\ell \right)  \right] \label{eq:dN1dv_window}\\
     f_\ell &=\begin{cases}
        \frac{1}{2} \left(1 - k \frac{v^c_\parallel}{v} \right) \text{ for }  v>v^c_\parallel \text{ where } \frac{1}{2}m \left(v^c_\parallel\right)^2 = V^\text{Al} + \mu B\\
        0 \hspace{43pt} \text{otherwise}\end{cases}\\
    f_h &=\begin{cases}
        \frac{1}{2} \left(1 - k \frac{v^c_\parallel}{v} \right) \text{ for } v>v^c_\parallel \text{ where } \frac{1}{2}m \left(v^c_\parallel\right)^2 = V^\text{Al} - \mu B \\ 
        \hspace{60pt} \text{and } V^\text{Al} > \mu B\\
        0 \hspace{40pt} \text{ for } v \leq v^c_\parallel \text{ where } \frac{1}{2}m \left(v^c_\parallel\right)^2 = V^\text{Al} - \mu B \\ 
        \hspace{60pt} \text{and } V^\text{Al} > \mu B\\
        \frac{1}{2} \hspace{40pt} \text{ for } V^\text{Al} \leq \mu B \label{eq:holdingCurve_windowEnd}  \end{cases}
\end{align}
This formalism enables a fit of both no-window and with-window UCN storage curves using the same set of free parameters.

\subsubsection{\label{subsubsec:holdingCurveFits}Multi-parameter fits of UCN storage curves}

Multi-parameter least square minimization fits were performed using the LMFIT Python library \cite{newville_matthew_2014_11813}. The results are presented in Tab.~\ref{tb:fitResults}. Data points were weighted using the total measurement uncertainty during the fitting process. For measurements with no window in the PM region, we fit Eqs.~(\ref{eq:holdingCurve_noWindowStart}) -- (\ref{eq:holdingCurve_noWindowEnd}), and for the measurements with a window present we fit Eqs.~(\ref{eq:holdingCurve_noWindowStart}) -- (\ref{eq:lossPerBounce}) and (\ref{eq:dN1dv_window}) -- (\ref{eq:holdingCurve_windowEnd}). It should be noted that the bottom of the prototype cell was located \qty{10}{\cm} above beam height. Therefore for NiP-coated Al wall run conditions we used an integration limits $v_\text{low}$ and $v_\text{high}$ as determined by $U=10$~\unit{\nano\eV} and $V^\text{guide} = 213$~\unit{\nano\eV} \cite{pattie_jr_evaluation_2017}. For dPS-coated PMMA wall run conditions we used $U=10$~\unit{\nano\eV} and $V^\text{cell} =160$~\unit{\nano\eV}.

For run condition 4,  we used a single effective loss per bounce parameter $\mu$ even though the cell was made of two different materials (i.e. dPS-coated cell wall with NiP electrodes). The reason for this was because the data did not have the sensitivity to resolve two separate $\mu$ values. This also applies to the simultaneous fit for data from run conditions 3 and 4.

\begin{table*}[ht]
\centering
\caption{\label{tb:fitResults}Multi-parameter fits of UCN storage curves. Refer to Tab. \ref{tb:runconditions} for run condition details.}
\begin{tabular}{
l
S
S
S
S
c
c}
\toprule
Run					& 1				& 2				& 3				& 4				& \multicolumn{2}{c}{3 \& 4}							\\
\midrule
$C/10^3$			& 2(1)			& 3(1)			& 5(3)			& 5(3)			& \multicolumn{1}{c}{$4(1)$} & $7(2)$					\\
$\gamma$			& 3.0(3)		& 2.5(2)		& 2.5(3)		& 2.7(2)		& \multicolumn{2}{c}{$\phantom{-}2.6(2)$}				\\
$\beta$				& 0.30(2)		& 0.8(2)		& 0.6(3)		& 0.8(2)		& \multicolumn{2}{c}{$\phantom{-}0.77(17)$}				\\
$k$					& 0.3(1)		& 0.6(1)		& 0.7(1)		& 0.5(1)		& \multicolumn{2}{c}{$\phantom{-}0.6(1)$}				\\
$\mu/10^{-4}$		& 2.0(1)		& 2.6(2)		& 3.5(4)		& 3.1(2)		& \multicolumn{2}{c}{$\phantom{-}3.4(2)$}							\\
Corr$(C,\gamma)$		& -1			& -0.94			& -0.96			& -0.96			& \multicolumn{1}{c}{$-0.97$}				& $-0.96$	\\
Corr$(C,\beta)$		& -0.2			& -0.32			& -0.4			& -0.5			& \multicolumn{1}{c}{$-0.49$}				& $-0.53$	\\
Corr$(C,k)$			& -0.7			& -0.14			& -0.14			& 0.02			& \multicolumn{1}{c}{$-0.16$}				& $-0.12$	\\
Corr$(C,\mu)$		& 0.9			& 0.85			& 0.77			& 0.79			& \multicolumn{1}{c}{$\phantom{-}0.78$}		& $\phantom{-}0.76$	\\
Corr$(\gamma,\beta)$	& -0.2			& 0.01			& 0.15			& 0.27			& \multicolumn{2}{c}{$\phantom{-}0.23$}					\\
Corr$(\gamma,k)$		& 0.71			& 0.44			& 0.37			& 0.22			& \multicolumn{2}{c}{$\phantom{-}0.41$}					\\
Corr$(\gamma,\mu)$	& -0.97			& -0.92			& -0.84			& -0.83			& \multicolumn{2}{c}{$-0.82$}							\\
Corr$(\beta,k)$		& -0.33			& -0.86			& -0.81			& -0.83			& \multicolumn{2}{c}{$-0.72$}							\\
Corr$(\beta,\mu)$	& 0.21			& 0.14			& 0.16			& -0.03			& \multicolumn{2}{c}{$\phantom{-}0.01$}					\\
Corr$(k,\mu)$		& -0.6			& -0.56			& -0.63			& -0.41			& \multicolumn{2}{c}{$-0.57$}							\\
Corr$(C_3,C_4)$		& 				& 				& 				& 				& \multicolumn{2}{c}{$\phantom{-}1\phantom{.00}$}		\\
\bottomrule
\end{tabular}
\end{table*}

\subsection{\label{subsec:PMramp}Spin dependent UCN rate as a function of the polarizing magnet field}

A flow-through measurement as described in Sec.~\ref{subsec:PMrampMeasurement} was taken to obtain information on the input UCN spectrum. The switcher was set to channel polarized UCN from the PM directly into the spin analyzer and detector. For the duration of the measurement, the strength of the PM was slowly decreased from 5~T to 0~T and the spin flipper was toggled on and off at regular intervals (as per Sec.~\ref{subsec:PMrampMeasurement}). The Al window was present in the PM region for this measurement.

\begin{figure}[htp]
    \centering
    \includegraphics[width=\columnwidth]{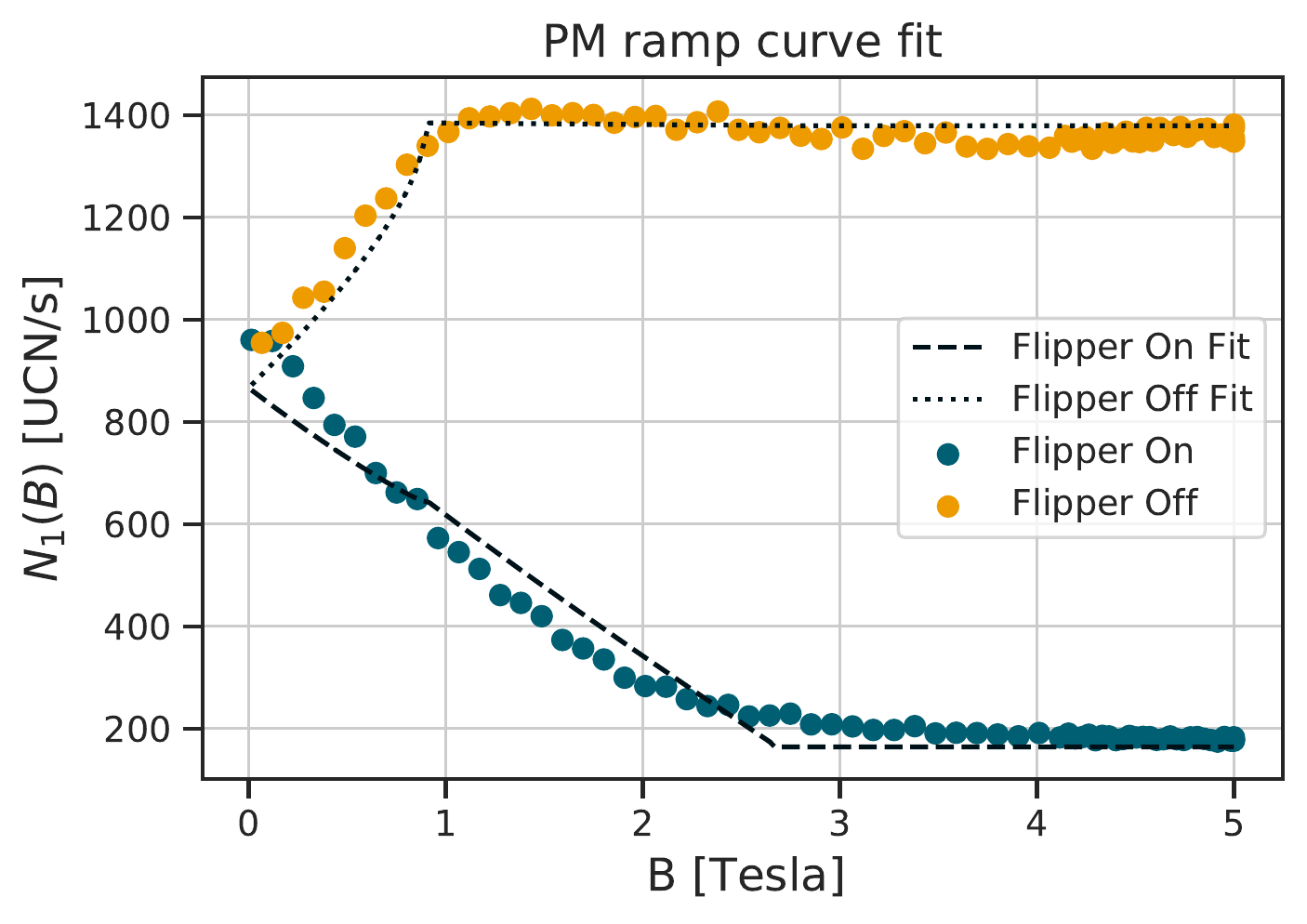}
    \caption{ A measurement of neutron transmission from the source to the drop-detector with the B field of the PM decreasing from 5 T to 0 T over the course of 72 minutes, with the spin flipper toggling on and off. This generates two curves that are describable using Eqs.~(\ref{eq:dN_off/dv}) and (\ref{eq:dN_on/dv}), which share the same set of free parameters as Eqs.~(\ref{eq:holdingCurve_noWindowStart}) -- (\ref{eq:lossPerBounce}) and  (\ref{eq:dN1dv_window}) -- (\ref{eq:holdingCurve_windowEnd}). The results of a simultaneous fit of these PM ramping curves are listed in Tab. \ref{tb:pmRampFit}}
    \label{fig:pmRampFit}
\end{figure}

The flow-through measurement provides two curves describing neutron rate for each spin state as a function of the PM field strength (see Fig.~\ref{fig:pmRampFit}). These curves can be fit using the formalism described in Sec.~\ref{subsubsec:holdingCurveFits} with modification to account for finite efficiency of the spin analyzer and spin flipper.

Dividing Eq.~(\ref{eq:dN1dv_window}) into expressions for low-field seekers and high-field seekers

\begin{align}
\frac{dN_{1,\ell}}{dv} &= \frac{C}{2}v^{\gamma} f_\ell = \frac{1}{2}v^{\gamma} f_{\ell D}\\
\frac{dN_{1,h}}{dv} &= \frac{C}{2}v^{\gamma} \left[ f_h + \beta \left( f_h - f_\ell \right)  \right] = \frac{1}{2}v^{\gamma} f_{h D}
\end{align}
where $f_{\ell D} = f_\ell$ and $f_{h D} = f_h +  \beta \left( f_h - f_\ell \right)$  describe the fraction of low-field seekers and high-field seekers downstream of the  PM, respectively.

For spin analyzing power $\epsilon_A$, when the spin flipper is off, there is some fraction of low-field seekers that are transmitted by the analyzing foil. Assuming that there is a negligible chance of depolarization downstream of the PM, the spectrum of the detected UCN with the spin flipper off is given by

\begin{gather}
    \frac{dN^\text{off}}{dv} = \frac{C}{2} v^{\gamma} \left[ \epsilon_A\, f_{h D} + (1-\epsilon_A)f_{\ell D}  \right]
    \label{eq:dN_off/dv}
\end{gather}

When the spin flipper is on, for spin flipping efficiency $\epsilon_S$, the spectrum of the detected UCN is

\begin{align}
    \frac{dN^\text{on}}{dv} &= \frac{C}{2} v^{\gamma} \big[ \epsilon_A\,\epsilon_S\, f_{\ell D} + \epsilon_A(1 - \epsilon_S) f_{h D} \nonumber \\
    &\quad + (1 - \epsilon_A)\epsilon_S\,f_{h D}  + (1 - \epsilon_A)(1-\epsilon_A)\,f_{\ell D} \big]\label{eq:dN_on/dv}
\end{align}

In the above equation, the first term describes low-field seekers downstream of the PM that are successfully spin flipped to high-field seekers and allowed through the spin analyzer. The second term is the fraction of high-field seekers that were not flipped and allowed through the spin analyzer. The third term is the fraction of high-field seekers that were flipped to low-field seekers, but allowed through to the detector due to finite analyzing power of the foil. The fourth term is the fraction of low-field seekers that were not spin flipped, but allowed through the spin analyzer due to finite analyzing power.

Following Ref.~\cite{ThorstenThesis}, we assume a spin analyzing power of $\epsilon_A=0.99$ for UCN in the energy range of the LANL nEDM experiment and a magnetized iron foil spin analyzer.

Spin flipping efficiency $\epsilon_S$ can be estimated as follows. With UCN in a flow-through mode from the source to the detector and with the PM at \qty{5}{\tesla}, when the spin flipper was turned on, detected UCN rate was reduced to $12\%$ compared to when the spin flipper was off, which corresponds to the ratio of Eq.~(\ref{eq:dN_on/dv}) to Eq.~(\ref{eq:dN_off/dv}). Assuming negligible depolarization downstream of the PM, this means $f_{\ell D} = 0$ and $f_{hD} = 1/2$. Letting $\epsilon_A=0.99$ and solving for $\epsilon_S$ yields a spin flipping efficiency of $\epsilon_S=0.89$.

The spin-flipper-off and spin-flipper-on curves were fit using an integral of $dN^{\text{off}}/dv$ and $dN^{\text{on}}/dv$, respectively, with limits of integration from \qtyrange{0}{7.58}{\meter\per\s} (the velocity equivalent of UCN in a \qty{5}{\tesla} field). This fit is shown in Fig.~\ref{fig:pmRampFit}, with the fit parameters from Tab. \ref{tb:pmRampFit}.

\begin{table*}
\centering
\caption{\label{tb:pmRampFit}Multi-parameter fit of a PM ramp}
\begin{tabular}{llllll}
\toprule
\multicolumn{1}{l}{Run condition} & \multicolumn{1}{l}{$\gamma$} & \multicolumn{1}{l}{$\beta$} & \multicolumn{1}{l}{$k$} \\ 
\midrule
With Al window, UCN-flow through mode           & 1.0(1)      & 0.00(5)                 & 0.30(2)   \\
\bottomrule
\end{tabular}
\end{table*}

\section{\label{sec:discussion}Discussion}

As measured in run condition 4, Fig.~\ref{fig:runCondition4_fit}, a storage time of \qty{180}{\s} in a single cell yields a UCN count of $\sim60,000$ UCN, well in excess of the number required to reach the desired statistical uncertainty. In Ref.~\cite{ito_performance_2018}, $\alpha=0.8$, $E=$~\qty{12}{\kilo\volt\per\cm}, $T_\text{fp}=$\qty{180}{\s}, $N=39,000$ for each cell, and $T_{\rm cycle} =$~\qty{300}{\s} were assumed to arrive at an estimated statistical uncertainty of $\sigma_{d_n}=4.0\times 10^{-26}e\cdot\text{cm}$ per day with a double cell geometry. With an assumed data taking efficiency of 50\% (for experiment down time, calibration and systematic studies) and the nominal LANSCE accelerator running schedule, this corresponds to a statistical uncertainty of $\sigma_{d_n}=2.1\times 10^{-27}e\cdot\text{cm}$ in 5 calendar years. The number of stored polarized UCN observed was about 50\% larger than what was reported in a similar measurement performed on the West Beamline~\cite{ito_performance_2018}. We attribute the improvement to the better switcher transmission in spite of the longer transport distance. 

In the envisioned LANL nEDM experiment, two precession chambers will need to be filled. We expect that the achievable stored UCN density in the precession chambers will be minimally affected by going from one cell, as was done in this commissioning experiment, to two cells. In general, the ultimate achievable density in a volume that is made of several sections is given by $\rho = P/(\Sigma_i V_i/\tau_i)$, where $P$ is the UCN production rate (or a rate at which UCN enter the volume), $V_i$ is the volume of a section in which UCN has a lifetime $\tau_i$.  The volume of our system is dominated by the volume of the guide that transports UCN from the source to the experiment. As a result, we estimate the reduction in the achievable density due to the additional volume of the second cell and the associated guide to be up to 10\%. 

Additionally, we have demonstrated the utility of a UCN storage curve model. Varying the strength of the PM for UCN storage curves allowed the properties of a continuous UCN input energy spectrum on the North Beamline to be obtained. Importantly, the model describes both window-in and window-out storage with the same input parameters. The model is able to describe the unintuitive feature of the window-in run condition curves, where as the PM field increases in strength the normalized counts also increase. This behavior was the primary motivation for the introduction of the free parameter $\beta$ during analysis. For small $\beta$ (closer to 0) the model predicts that the number of UCN decreases as PM strength increases, but for larger $\beta$, low field seekers are spin-flipped and may pass through the PM region as a high field seeker, resulting in the behavior observed in Fig.~\ref{fig:runCondition2_fit} -- \ref{fig:runCondition4_fit}.

We can draw a number of observations from the results of the UCN storage curves fits depicted in Tab.~\ref{tb:fitResults}. First, the value of $C$ in run conditions 3 and 4 is larger than the value in run condition 2, showing that the new rotary switcher transports a larger number of UCN than the old prototype switcher. Second, $k$ (ranging from 0.3(1) to 0.7(1)) across all run conditions indicates that the angular distribution of the input UCN velocity spectrum on the beamline is relatively forward directed. Third, $\beta$ is $\sim 0.7$ for run conditions with the Al window and is $\sim 0.3$ for the run condition without the window. This is consistent with the idea that the presence of the Al window reduces the number of low-field seeking UCN that pass the PM region, resulting in an increased chance for a low-field seeker to see a depolarizing region upstream of the PM and an effectively larger $\beta$. Comparatively, in the no window case, more low-field seekers pass the PM region, lowering the chance for a low-field seeker to be spin-flipped and giving a smaller value for $\beta$. 

In some cases there are rather large correlations among parameters as seen in Tab.~\ref{tb:fitResults}. The largest parameter correlations for fits across all run conditions are  $\text{Corr}(C,\gamma) \sim -1$, $\text{Corr}(\gamma,\mu) \sim -0.9$, and $\text{Corr}(C,\mu) \sim 0.8$. Special care was taken in the least-squares fitting process to choose initial parameter guesses for $\gamma$ and $\mu$ that were based on previous measurements along the West Beamline (see Ref.~\cite{ito_performance_2018, pattie_jr_evaluation_2017}) because of these correlations. As a result, we obtained $\gamma \sim 3$ and $\mu \sim 3\times 10^{-4}$ across all run conditions, where $\gamma$ was constrained to the range $[1,5]$ and $\mu$ was constrained to the range $[10^{-3}, 10^{-5}]$ during fitting. These values are consistent with previously measured values of $\mu = 1.4(2)\times 10^{-4}$ (Ref.~\cite{pattie_jr_evaluation_2017}) and $\gamma = 2.9(0.6)$ (Ref.~\cite{ito_performance_2018}).

It can be seen in Fig.~\ref{fig:runCondition1_fit} that the storage curve model reproduces the storage curves at each PM field strengths for configurations without a window. For configurations with a window in the PM region (Fig.~\ref{fig:runCondition2_fit} -- \ref{fig:runCondition4_fit}), the holding curve model reproduces the storage curves for holding times less than \qty{200}{\s}. For longer holding times ($>$~\qty{200}{\s}) in run conditions with the PM window, the holding curve model overpredicts the number of UCN. This is especially apparent in Fig.~\ref{fig:runCondition2_fit}, for holding times of \qty{500}{\s} for $B=3.75\text{ T}$ and $B=2.5\text{ T}$, as well as the $B=0\text{ T}$ and $B=1.25\text{ T}$ curves for holding times greater than \qty{200}{\s}. 

The statistical and leak correction uncertainties for longer holding time data points are too small to attribute the overprediction purely to weighting of the data points during the fitting process. In addition, the model is able to describe the behavior of the curves in run condition 1 for all PM field strengths.  We postulate that there is an additional mechanism caused by the presence of the window that preferentially selects for higher energy UCN to enter the precession cell. Higher energy UCN experience more wall collisions within the precession cell and have a higher loss per bounce, increasing the chance to be lost during the storage period.

One possible mechanism is bulk scattering of UCN within the PM window caused by nonuniformities within the window material. The magnetic field accelerates high-field seeking UCN through the Al window. If the window scatters the UCN, then a fraction of the longitudinal velocity (i.e. along the direction of the beamline) results in a transverse velocity component. Lower energy UCN may lose enough longitudinal velocity and become trapped in the the PM field potential well, a process less likely for higher energy UCN.

Lastly, we have shown that elements of the holding curve model may be adapted to a UCN flow-through measurement through a PM with an Al window, while the PM slowly ramps up or down. We observe that the model is able to reproduce the major features of the data (Fig.~\ref{fig:pmRampFit}). As the B field increases, low-field seekers are reflected by the potential well of the PM. Conversely, the fraction of high field seekers allowed past the Al window slowly increases with the strength of the B field, until the strength of the B field reaches a threshold where all high-field seekers are accelerated past the window region. The parameters of the model, presented in Tab. \ref{tb:pmRampFit}, indicate a forward-directed angular distribution ($k=0.30(2)$) consistent with the holding curve fits in Tab. \ref{tb:fitResults}. The estimate of $\beta=0$ is also consistent with what we expect, because UCN in a flow-through measurement with no preload period have less opportunity to be spin-flipped upstream of the PM.

The precision for $\gamma$ in Tab. \ref{tb:pmRampFit} is limited due the change in the UCN energy spectrum during the course of the flow-through measurement. During the \qty{72}{\minute} period required to ramp the PM, the continuous use of the UCN source causes heating and a corresponding change in the input energy spectrum \cite{anghel_solid_2018}. Note that the measurement depicted in Fig.~\ref{fig:pmRampFit} was a PM ramp down sequence, where the field was decreased from \qtyrange{5}{0}{\tesla}. Since there is no time dependence for $\gamma$, the model is less descriptive of the data obtained near the end of the ramp (\qtyrange{1}{0}{\tesla}).

\section{\label{sec:summary}Conclusion}

We have demonstrated successful instrumentation of a prototype precession cell, single-channel spin analyzer, spin flipper, UCN detector, and rotary switcher for UCN transport in preparation for the LANL nEDM experiment. Approximately 60,000 UCN, sufficient to obtain a statistical uncertainty of $\delta d_n = 2\times 10^{-27}$~$e\cdot\text{cm}$ for the nEDM have been stored in a prototype cell with dPS-coated PMMA walls and NiP-coated Al electrode plates. An analytical model has been developed that describes various properties of the UCN exiting the PM that affect the performance of experiments that use UCN. These properties include depolarization in the beamline upstream of the PM, the UCN velocity spectrum, and the angular distribution of the UCN. We have successfully applied this model to the analysis of a series of UCN storage curves taken with prototype precession cells of the LANL nEDM experiment and data on UCN transmission through the PM. The extracted UCN velocity spectrum can be used to analyze and optimize the performance of the LANL nEDM experiment. 

\section*{Acknowledgments}
This work was supported by Los Alamos National Laboratory LDRD Program (Project No. 20190041DR), the U.S. National Science Foundation (Grant No. PHY-1828512 (Indiana U.) and Grant No. PHY-2110988 (U. Michigan)), and the U.S. Department of Energy, Office of Science, Office of Workforce Development for Teachers and Scientists, Office of Science Graduate Student Research (SCGSR) program (Contract No. DE‐SC0014664). We gratefully acknowledge the support provided by the LANL Physics and AOT Divisions.


\bibliographystyle{elsarticle-num-names}
\bibliography{refs}

\end{document}